\documentclass[]{elsart}

\usepackage[dvips]{graphicx}
\usepackage{amsmath}
\usepackage{amssymb}

\newcommand{\erw}[1]{\langle #1 \rangle}
\newcommand{\ca}[1]{ {\mathcal{#1}}}

\begin{document}

\begin{frontmatter}

\title{A master equation approach to option pricing}
\author[a]{D. Faller}
\ead{\\Daniel.Faller@physik.uni-freiburg.de}
\author[a,b]{F. Petruccione}
\ead{\\Petruccione@physik.uni-freiburg.de}
\address[a]{Albert-Ludwigs-Universit\"at, Fakult\"at f\"ur Physik, \\
         Hermann-Herder Stra{\ss}e 3, D--79104 Freiburg im Breisgau,
         Federal Republic of Germany}
\address[b]{Istituto Italiano per gli Studi Filosofici, Palazzo Serra di Cassano,\\
           Via Monte di Dio 14, 80132 Napoli, Italy}

\date{\today}


 \begin{abstract}
A master equation approach  to the numerical solution of option
pricing models is developed. The basic idea of the approach is to
consider the Black--Scholes equation as the macroscopic equation of an
underlying mesoscopic stochastic option price variable. The dynamics
of the latter is constructed and formulated in terms of a master
equation. The numerical efficiency of the approach is demonstrated by
means of stochastic simulation of the mesoscopic process for both
European and American options.
\end{abstract}

\end{frontmatter}

\section{Introduction}

As is well-known the seminal
Black-Scholes analysis \cite{black_scholes} which leads to the fair value of an option is based 
mainly upon the following assumptions. First the stock price, i.e. the underlying, 
follows a geometric Brownian motion. Second a hedge position is formed with a portfolio
of short underlying and a long position of a number of European
options. Then an arbitrage argument leads to the renowned Black--Scholes
partial differential equation determining the value of the option.
Of course, for simple cases, e.g. constant interest rate and volatility,
explicit analytical solutions to the equation are known \cite{hull.1997}. 
However, for more involved cases one has to rely upon numerical methods. 
Here we are not concerned with deterministic methods, e.g. finite differences,
but rather with Monte Carlo methods \cite{dupire}.\\
The idea behind the canonical Monte Carlo method is to exploit the fact
that the fair value of an option is given by the present value of the expected
payoff at expiry under the risk neutral measure. Thus the standard
Monte Carlo approach is based upon the simulation of a geometric
Brownian motion for the underlying asset until expiry. Then the payoff is
computed and discounted up to the current time. By averaging over
different realizations of this stochastic process the current option price is
estimated \cite{boyle.1977}. In spirit this approach is similar to
the standard Monte Carlo random walk technique for the solution of
partial differential equations with boundary value problems \cite{sabelfeld}.\\
Monte Carlo methods offer easy to understand numerical algorithms
which can be easily applied to quite complicated, e.g. path
dependent or correlated multi-asset, options \cite{wil1}. In addition they allow for a
straightforward inclusion of stochastic terms such as
the interest rate or volatility. From a numerical point of view they are
especially suited to problems with many degrees of freedom and the
algorithm can be easily run in parallel. \\
However Monte Carlo simulations of the Black-Scholes equation are
usually slower than comparable finite difference solutions of the
partial differential equation.  Another disadvantage of the standard
Monte Carlo approach is that while the application to European options
is straightforward, the valuation of American options is more
involved. Using a generalisation of the canonical Monte Carlo method
one has to assure that the early exercise condition stating that the
option price is always above the current payoff is not
violated sometime in the future. If this happens and the option
value is below the payoff, the option is exercised and the option value is
given by the payoff function. A Monte Carlo simulation based on the
dynamics of the underlying asset has to keep track of all these points in asset time
space  which  makes the algorithm  ineffective.
Nevertheless more advanced Monte Carlo algorithms are available
\cite{longstaff2001,broadie1997a,bossaerts1989,boyle1997}.\\
Since the Monte Carlo methods described above are based on a model of the
underlying dynamics of asset values these approaches could be named
microscopic. 
In contrast now we are going to follow a different
strategy, which we would like to name  
mesoscopic, which has already been applied with success to simulations of 
turbulence in fluid dynamics \cite{hpb5,hpb6,hpb7}, to the
investigation of hydrodynamic fluctuations \cite{hpb8,hpb4} as well as
to magnetohydrodynamics \cite{grauer.1996}. The same approach has been
shown to lead to fast Monte Carlo algorithms for the balance equations
of nonequilibrium thermodynamics \cite{hpb3}, for the simulation of
chemical reactions \cite{hpb11} and reaction--diffusion processes \cite{hpb12}.\\
The basic idea of the mesoscopic approach \cite{kampen.1992} is to regard the value of the 
option, say $V$ as the expectation value of an underlying fluctuating 
(mesoscopic) stochastic option value, say $\theta$. 
A master equation for $\theta$ is easily constructed in such a way that 
the expectation value of the multivariate stochastic variable 
$\theta$ satisfies the macroscopic Black--Scholes equation. The 
stochastic process defined in terms of a master equation is easily 
simulated and allows for a  Monte Carlo algorithm for the direct 
simulation of the Black--Scholes equation.\\
This paper is structured as follows. In section
\ref{sec:black-schol-equat} the mesoscopic master equation approach is
motivated. The dynamics of the underlying asset is modelled as a
piecewise deterministic process (PDP). The derivation of the fair
option price for a mesoscopic stochastic option value, which parallels
the Black--Scholes one, shows that the expectation value of the
stochastic option price satisfies the ``macroscopic'' Black--Scholes
equation. 
In section \ref{sec:mast-equat-form} a master equation for the mesoscopic fair
option value is constructed, and a stochastic simulation
algorithm is applied to the valuation of European and American
options. In the last section perspectives and further development of
the approach are indicated.

\section{Black--Scholes Equation}\label{sec:black-schol-equat}

The standard textbook derivation of the Black--Scholes equation
assumes a geometric Brownian motion for the underlying
asset which is described by the stochastic differential
equation
\begin{equation}\label{eq:ds}
  ds = \mu s \, dt + \sigma s \, d W\;.
\end{equation}
Here $s$ is the asset value, $\mu s$ and $\sigma^2 s^2$ denote drift
and variance of the random walk. Since $d W$ is the increment of a Wiener process the
resulting trajectory will be continuous. For  more realistic models
instantaneous jumps of the underlying, described by
additional stochastic terms based on Poisson
processes \cite{merton,cox}, may be included in (\ref{eq:ds}). These
so called jump diffusion models, are used e.g. to
incorporate the effect of information about stocks which arrives at
random times \cite{wil1}.

\subsection{Black--Scholes equation from a piecewise deterministic
  process (PDP) \label{sec:black-schol-poisson}}

In contrast to the standard derivation of the Black--Scholes equation
based on 
(\ref{eq:ds}),
this section demonstrates how 
the derivation can be paralleled by modelling the underlying in
terms of a piecewise deterministic process \cite{davis.1993}.
This is possible since a diffusion process can be
represented as the continuous limit of  an appropriate jump process.   
This approach, despite describing the same dynamics as the
standard approach, has the advantage of clearly demonstrating that it
is quite obvious to interpret the option value as a stochastic
variable and thus to interpret the Black--Scholes equation as the
macroscopic expectation value of a mesoscopic stochastic option value.
In addition the 
formulation in terms of PDPs 
simplifies the inclusion of additional jump processes since one has to
deal only with one type of stochastic process. Another important point
is that for such processes one can easily proceed to a master
equation formulation, for which
powerful numerical algorithms exist, as will be shown later on.\\
Consider now the stochastic differential equation 
\begin{equation}\label{eq:d_x}
ds = s \mu \, dt + \delta \! s \, dN^+ - \delta \! s \, dN^-\;,
\end{equation}
with $\erw{dN^{\pm}}= s^2 \sigma^2/(2 \delta \! s^2)\, dt$. The
Poisson increments $dN^{\pm}$ are either $0$, in which case the
deterministic evolution takes place, or $1$, which describes an
instantaneous jump of size $\pm \delta \! s$. 
Stochastic process of this general form are thus also known  as
piecewise deterministic processes  \cite{hp_frpe_book}. \\
In the limit
$\delta \! s \rightarrow 0$ this stochastic differential equation
corresponds to  a geometric Brownian motion
with drift $s \mu$ and variance $s^2 \sigma^2$ \cite{gardiner}. Hence in this limit the
above equation generates the same dynamics as Eq. (\ref{eq:ds}), but
is  formulated as a PDP.\\ 
Using the above stochastic differential Eq. (\ref{eq:d_x}) instead of
Eq. (\ref{eq:ds}) as starting point for the classical Black--Scholes
derivation \cite{wil1} one obtains in the limit $\delta \! s \rightarrow 0$ for
the expectation value of the stochastic option price $v$
\begin{eqnarray}
\erw{\frac{\partial v(s,t)}{\partial t}} = 
 - \frac{s^2 \sigma^2}{2} \erw{\frac{\partial^2
    v(s,t)}{\partial s^2}}
 - r s \erw{\frac{\partial
    v(s,t)}{\partial s}}  + r \erw{v(s,t)} \;,
\end{eqnarray}
where $r$ denotes the riskfree interest rate.
If the expectation value $V(S,t)=\erw{v(s,t)}$ is identified with the
macroscopic option price then one obtains the Black--Scholes
equation for a European option on a non dividend paying asset.\\
Hence the traditional Black-Scholes equation can be interpreted as the expectation
value of a stochastic process. Additional not infinitesimal jump processes can
now easily be included, but this of course will change the hedging
strategy \cite{merton,cox} and the expectation value of the option
price will then follow equations similar to the Black--Scholes equation.

\subsection{Macroscopic Black--Scholes equation}

The above derivation assumed a non-dividend paying stock, this
assumption can be easily omitted to obtain \cite{wil1,wil2}
\begin{eqnarray}\label{eq:bs}
\frac{\partial\,  V(S,t)}{\partial t}  =  - \frac{\sigma^2 S^2 }{2} \,
\frac{\partial^2\,  V(S,t)}{\partial S^2} - (r-q) \, S \frac{\partial \, V(S,t)
  }{\partial S}  + r\, V(S,t)\;.
\end{eqnarray}
Here  $\sigma$ and $r$ again denote the volatility and the interest
rate and the continuous dividend is given by $q$. 
In order to numerically solve this partial differential equation the
relevant variables are made dimensionless and the time direction is reversed, i.e. we have,
\begin{equation}\label{eq:transformations}
\tilde S = \ln(S/E), \;\;     {\tilde
  V}(\tilde S,\tau) = V(S,t)/E, \;\; \tau = \frac{\sigma^2}{2}(T-t).
\end{equation}
In this way one obtains
\begin{eqnarray}\label{eq:bs_dl}
\frac{\partial\,  \tilde V(\tilde S,\tau)}{\partial \tau}  =
\frac{\partial^2\,  \tilde V(\tilde S,\tau)}{\partial \tilde S^2}  +
(k_q-1) \frac{\partial  \tilde V(\tilde S,\tau)
  }{\partial \tilde S}  - k_r\, \tilde V(\tilde S,\tau)\, ,
\end{eqnarray}
where $k_q=2\,(r-q)/\sigma^2$ and $k_r=2\,r/\sigma^2$ have been
introduced.
If the parameters of the Black--Scholes equation~(\ref{eq:bs}) are
assumed to be constant in time, then the analytic solution to the
above equation is available, see e.g. \cite{wil1}.

\section{Master Equation formulation of the Black-Scholes Equation}
\label{sec:mast-equat-form}

As seen in section \ref{sec:black-schol-poisson} the Black--Scholes
equation can be interpreted as a deterministic equation which governs
the expectation value of a stochastic option price. In this section we
will describe how the stochastic dynamics of the option price can be
formulated in terms of a master equation.
 For the sake of simplicity we will consider the Black-Scholes
equation for  a European call in the dimensionless form of
Eq.~(\ref{eq:bs_dl}). The generalisation to a put
is straightforward and the application to American options is
discussed in section \ref{sec:american-options}.

\subsection{General Theory}\label{sec:general-theory}

A general approach in statistical physics  is to derive macroscopic
equations from a known microscopic dynamics. Here 
the opposite approach is followed: given a partial differential equation for a
macroscopic observable we
construct a mesoscopic stochastic process such that the 
expectation value of the stochastic variable 
is governed by the original partial differential equation. In this section
the general theory needed for the construction of a stochastic process
underlying a partial differential equation is presented, for a review see \cite{hpb1}.\\
In order to define the underlying  stochastic process the
state space of the system has to be given. To this end the space
variable $x$ is discretized such that $\theta_\lambda$ denotes the state of the system at the
discrete position $x_\lambda, \;\; \lambda = 1, \dots ,n$. Hence the
state space $\Gamma$ of the system is given by $\Gamma = \{\theta \mid
\theta \in {\mathbb R}^n\}$.
The stochastic
 process is defined by the joint probability distribution $P =
 P(\left\{\theta_\lambda\right\},t)$ giving the probability of finding
 the values $\{\theta_\lambda\}$ at time $t$. The time development of
 the probability density $P$ is given by a master equation of the
 general form
\begin{equation}\label{eq:evolution_p}
\frac{\partial }{\partial t} P(\theta,t) = {\ca{A}} P(\theta).
\end{equation}
The operator $\ca{A}$ is defined in terms of diffeomorphic
maps $b$ acting on the state space $b: \Gamma \rightarrow \Gamma$, and
corresponding operators ${\ca B}_b$ acting on 
functions $F:  \Gamma \rightarrow {\mathbb R}$ in
the following way
\begin{equation} \label{eq:def}
\ca{B}_b F(\theta) = F(b^{-1}(\theta)).
\end{equation}\\
The most general operator  $\ca{A}$ needed in this paper is of the form
\begin{equation} \label{eq:general.a}
\ca{A} = c \sum_\mu [\text{Det}(b_\mu)]^{-1} \ca{B}_{b_\mu} - {\mathbb{I}},
\end{equation}
where $\text{Det}(b_\mu)$ denotes the determinant of the Jacobian matrix of
the map $b_\mu$ and ${\mathbb I}$ denotes the identity. In order to generate a valid stochastic process
the operator $\ca{A}$ has to fulfil (i)
all transition probabilities $w$ are positive and (ii)
$\erw{\ca{A} F} = 0$ in order to preserve the normalisation
of the probability density.\\
With these definitions the macroscopic equation of motion for a general
observable $F$ can now easily be computed, one obtains
\begin{equation}\label{eq:real_expectation_formula}
\frac{\partial}{\partial t} \erw{F} = c
\sum_\mu \erw{F(b_\mu(\theta)) -F(\theta)}.
\end{equation}
The above equation enables the computation of the macroscopic
expectation value of a general observable $F$ given an arbitrary
multivariate Markov process defined by the time evolution operator
$\ca{A}$.\\
For a master equation formulation of the stochastic process one needs
the transition probability 
\begin{equation}
w(\theta, \tilde \theta) = c \sum_\mu \delta(\theta, b_\mu(\tilde \theta)).
\end{equation}
Hence the canonical form of the master equation is obtained
\begin{equation}
\frac{\partial}{\partial t} P(\theta,t)= \int D \tilde \theta \left\{ w(\theta, \tilde \theta) P(\tilde \theta,t) - w(\tilde \theta, \theta) P(\theta,t)\right\}.
\end{equation}
Now the approach sketched above is applied to the Black-Scholes
equation. Hence the interesting price range of the underlying $x= \tilde S$ is divided in $n$ discrete
points $x_\lambda,\, \lambda=1, \dots ,n$ with distances 
$\Delta_\lambda=x_\lambda-x_{\lambda-1},\, \lambda=2, \dots, n$.

\subsection{Stochastic process with uniform discretization}

The easiest possibility to construct the stochastic process is to use
a uniform discretization of the state space $\Delta_\lambda = \delta
l$. The Black-Scholes equation (\ref{eq:bs_dl}) consists of a diffusive,
a convective and a part corresponding to a chemical reaction. The
stochastic process underlying the Black-Scholes equation will also
consist of three parts corresponding to these terms in the partial
differential equation. The
total time evolution operator $\ca{A}$ is thus defined by
\begin{equation}\label{eq:bs_a_op_uniform_grid}
\ca{A}_{\text{BS}} = \ca{A}_{\text{diff}} + \ca{A}_{\text{conv}} +
\ca{A}_{\text{chem}}.
\end{equation}
The above time evolution operators are defined through their
corresponding maps, see Eq.  (\ref{eq:general.a}). 
The map 
\begin{equation}\label{eq:a_mu}
  a^{\pm}_\mu = \begin{cases}
    \theta_\mu &\rightarrow \theta_\mu - \alpha \, \theta_\mu  \\
    \theta_{\mu \pm 1} &\rightarrow \theta_{\mu \pm 1} + \alpha \, \theta_\mu
    \\
  \end{cases},
\end{equation}
describes the diffusive part, while the maps $b_\mu$ and $c_\mu$
\begin{equation}\label{eq:b_mu}
  b_\mu = \begin{cases}
    \theta_\mu &\rightarrow \theta_\mu - \alpha \, \theta_\mu  \\
    \theta_{\mu-1} &\rightarrow \theta_{\mu -1} + \alpha \, \theta_\mu
    \\
  \end{cases},\qquad
  c_\mu = \begin{cases}
    \theta_\mu \rightarrow \theta_\mu - \alpha \, \theta_\mu  
  \end{cases},
\end{equation}   
correspond to the convective part and to the chemical reaction term. 
Hence the time evolution operators for the three parts of the stochastic process
are given by
\begin{eqnarray}\label{eq:process_simple}
\ca{A}_{\text{diff}} &=& \frac{1}{\alpha \delta l^2} \sum_\mu
\frac{1}{1-\alpha} \{ \ca{A}_\mu^+ + \ca{A}_\mu^- \} -2 \,{\mathbb  I},\\
\ca{A}_{\text{conv}} &=& \frac{k-1}{\alpha \, \delta l} \; \sum_\mu \frac{1}{1-\alpha} 
\ca{B}_\mu - {\mathbb I},\\
\ca{A}_{\text{chem}} &=&  \frac{k}{\alpha} \; \sum_\mu
\frac{1}{1-\alpha} \ca{C}_\mu - {\mathbb I}.
\end{eqnarray}
Here the operators $\ca{A}_\mu^\pm, \ca{B}_\mu, \ca{C}_\mu$ are
defined according to the general definition (\ref{eq:def}) through their maps 
$a_\mu^\pm, b_\mu, c_\mu$. In order to prove that the expectation
value of the stochastic process whose generator is given by
(\ref{eq:bs_a_op_uniform_grid})  really solves the Black-Scholes equation one uses
Eq. (\ref{eq:real_expectation_formula}) and the projection operator
$F_\lambda(\theta)= \theta_\lambda$. One then immediately obtains
\begin{eqnarray}
 \frac{\partial}{\partial t} \erw{\theta_\lambda} =
\frac{\erw{\theta_{\lambda+1}} + \erw{\theta_{\lambda-1}} -  2
  \erw{\theta_{\lambda}}}{\delta l^2} 
+ (k-1) \frac{\erw{\theta_{\lambda+1}} - \erw{\theta_\lambda}
}{\delta l} - k \erw{\theta_\lambda},
\end{eqnarray}
which in the continuum limit $\delta l \rightarrow 0$ converges
towards the dimensionless Black-Scholes equation (\ref{eq:bs_dl}). \\
With these definitions the transition probability 
$w(\theta, \tilde \theta)$ becomes
\begin{eqnarray}
w(\theta, \tilde \theta) &=& \frac{1}{\alpha \delta l^2} \sum_\mu 
\delta(\theta, a_\mu^+(\tilde \theta)) + \delta(\theta, a_\mu^-(\tilde
\theta)) +\frac{k-1}{\alpha \delta l} \sum_\mu \delta(\theta, b(\tilde
\theta)) \nonumber\\ && + 
 \frac{k}{\alpha} \sum_\mu \delta(\theta, c(\tilde \theta)).
\end{eqnarray}
Summarising the Black-Scholes Eq. (\ref{eq:bs_dl}) can be numerically solved by
simulating the  stochastic process
\begin{eqnarray}\label{eq:process_uniform_grid}
\frac{\partial}{\partial t} P(\theta,t) &=& \ca{A}_{\text{BS}}
P(\theta) 
= \int D \tilde \theta \left\{ w(\theta, \tilde \theta) P(\tilde
  \theta,t) - w(\tilde \theta, \theta) P(\theta,t)\right\}.
\end{eqnarray} 
\begin{figure}
\begin{center}
\includegraphics[width=0.75\linewidth]{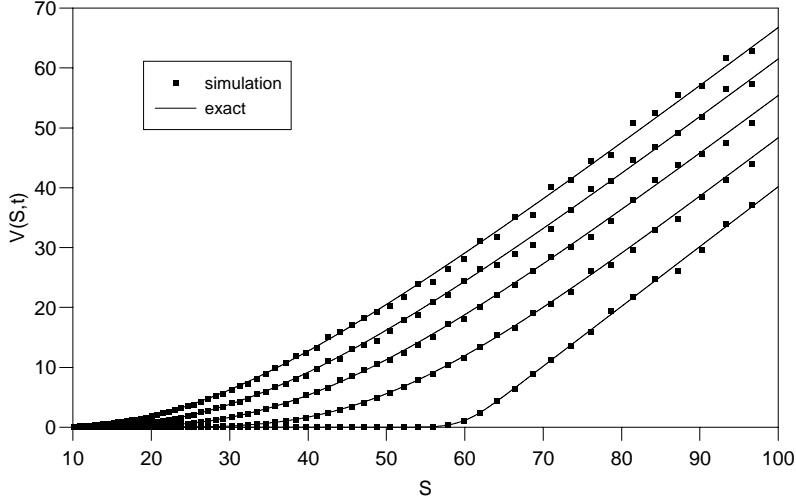}
\caption{Time evolution of the analytic solution (continuous line) and results of a direct stochastic
  simulation (squares) according to Eq.~(\ref{eq:bs_a_op_uniform_grid}) of the
  Black-Scholes equation for an European call with
  Exercise price 60, $\sigma=0.2$, $r=0.06$ for five different points
  in time. The solution of the master equation was estimated by
  averaging 10 realization with $\alpha=0.005$ at times $T=10, 7.5, 5,
  2.5, 0.05$.
\label{fig:fig_2}}
\end{center}
\end{figure}
This is usually
done with the help of the following algorithm:
\begin{itemize}
\item initialise $\theta_\lambda$ at $t=0$
\item while $t < t_{\text{end}}$
  \begin{itemize}
  \item compute random time step $\tau$ until the next jump occurs
    from an exponential distribution with mean value
    $\erw{\tau}=1/w_{\text{tot}}$ \item apply one of the transitions 
    $a^\pm_\mu, b_\mu, c_\mu$ selected
    according to their relative probability
  \end{itemize}
\end{itemize}
By repeating the above algorithm different realizations of $\theta$ are generated
and thus $\erw{\theta_\lambda}$ can be estimated from a sample of
realizations. \\
This stochastic process has some interesting features. 
The parameter $\alpha$ can be used to control the size of the fluctuations. 
In earlier applications of the general theory
presented above to balance equations of non-equilibrium statistical
mechanics, the parameter $\alpha$ could be 
interpreted as the temperature of the physical system
\cite{hpb1,hpb2}. 
The larger the parameter $\alpha$ the larger are the fluctuations, 
hence it is intuitively clear that,
e.g. in a thermodynamic setting $\alpha$ plays essentially the role of the
temperature.\\
From a numerical point of view the total transition probability
$w_{\text{tot}}$ is constant. Thus the error made by taking a constant time step
$\tau=1/w_{\text{tot}}$ in the numerical simulation vanishes as $O(\delta t)$. This significantly reduces the
number of random variables needed.
In addition the transition probabilities do not depend on the current state of the
system. This makes the random selection of a transition very
efficient, otherwise algorithms as discussed in \cite{fricke1} have
to be used.\\
Fig. \ref{fig:fig_2} shows a typical result of a
simulation averaged over $10$ realizations of the stochastic process
with $\alpha=0.005$.  Naturally the  simulation results fluctuate around the
analytic solution, but these fluctuations are very small if one takes
into account that  only $10$ realizations are averaged. The reason for
this of course is the parameter $\alpha$. The smaller $\alpha$, the
smaller are the fluctuations of the process and the smaller is the number
of realizations needed. \\
One drawback of this stochastic process is, that it is restricted to
a uniform discretisation of the dimensionless variable $\tilde S$ from
Eq. (\ref{eq:bs_dl}). As can be seen in Fig. \ref{fig:fig_2} after
transforming back to the original variables $V$ and $S$ this leads
to an exponential distribution of the discrete values of the
underlying, see Eqs.~(\ref{eq:transformations}).
Hence most of the points are at the left border of the integration
region. 
Since the total transition
rate performs as $1/\delta l^3$ the length of the time step during the
simulation and thus the overall computing time depends strongly on
the discretisation of the underlying. A numerically more
efficient algorithm thus has to use a uniform discretization of the
underlying $S$ which hence requires much less points. This results in a
non-uniform discretization of the dimensionless variable $\tilde S$ in
Eq. (\ref{eq:bs_dl}).

\subsection{Stochastic process with non-uniform discretization}

To enable a non-uniform discretization of the Black-Scholes equation (\ref{eq:bs_dl})
one uses an arbitrary distance $\Delta_\lambda$ of two neighbouring
points. The time evolution operator $\ca{A}_{\text{chem}}$ from the
previous section does not depend on the discretization and is thus not
altered. But for the diffusive and convective part of the
Black-Scholes equation new stochastic processes
$\ca{A}^{\text{n.u.}}_{\text{diff}}$ and
$\ca{A}^{\text{n.u.}}_{\text{conv}}$ taking into account the
non-uniform (n.u.) discretisation have to be defined.
With the definitions
\begin{eqnarray}\label{eq:r_mu}
 \!\!\! \!\!\!\! \!  r^{+}_\mu \! \! &=& \! \!\begin{cases}
    \theta_\mu  \! \!\! \!&\rightarrow \theta_\mu - \alpha_1 \,
    \frac{\theta_\mu}{\Delta_{\mu-1} \Delta_\mu} \\
    \theta_{\mu + 1}  \! \!\! \!&\rightarrow \theta_{\mu + 1} + \alpha_1 \, \frac{2
      \theta_\mu \Delta_{\mu+1}}{(\Delta_{\mu+1}+\Delta_\mu) \Delta_{\mu} \Delta_{\mu+1}}
    \\
  \end{cases},\\
\!\! \!\!\! \!\! \! r^{-}_\mu \! \! &=& \! \!\begin{cases}
    \theta_\mu  \! \!\! \!&\rightarrow \theta_\mu - \alpha_1 \,
    \frac{\theta_\mu}{\Delta_{\mu-1} \Delta_\mu} \\
    \theta_{\mu - 1}  \! \!\! \!&\rightarrow \theta_{\mu - 1} + \alpha_1 \, \frac{2
      \theta_\mu \Delta_{\mu-2}}{(\Delta_{\mu-1}+\Delta_{\mu-2}) \Delta_{\mu-2} \Delta_{\mu-1}}
    \\
  \end{cases}\!\!,
\end{eqnarray}
for the diffusive and 
\begin{eqnarray}\label{eq:s_mu}
  s_\mu &=& \begin{cases}
    \theta_\mu &\rightarrow \theta_\mu - \alpha_2 \, \frac{\theta_\mu}{\Delta_{\mu}}  \\
    \theta_{\mu-1} &\rightarrow \theta_{\mu -1} + \alpha_2 \, \frac{\theta_\mu}{\Delta_{\mu-1}}
    \\
  \end{cases},
\end{eqnarray}
for the convective part one obtains  the two new time evolution operators
\begin{eqnarray}\label{eq:process_adv}
\!\!\!\!\!\!\!\!\ca{A}^{\text{n.u.}}_{\text{diff}}\!\! &=&\!\! \frac{1}{\alpha_1 } \sum_\mu
\frac{1}{\text{Det}(r^+_\mu)}  \ca{R}_\mu^+ +
\frac{1}{\text{Det}(r^-_\mu)}\ca{R}_\mu^-  -2 \, {\mathbb I},\\
\!\!\!\!\!\!\!\!\ca{A}^{\text{n.u.}}_{\text{conv}}\!\! &=&\!\!
\frac{k-1}{\alpha_2} \; \sum_\mu \frac{1}{\text{Det}(s_\mu)}  
\ca{S}_\mu - {\mathbb I}.
\end{eqnarray}
The total time evolution operator $\ca{A}^{\text{n.u.}}_{\text{BS}}$ is now given by
\begin{equation}\label{eq:a_op_non_uniform}
\ca{A}^{\text{n.u.}}_{\text{BS}} = \ca{A}^{\text{n.u.}}_{\text{diff}} +
\ca{A}^{\text{n.u.}}_{\text{conv}} + \ca{A}_{\text{chem}}.
\end{equation}
In order to prove that the expectation value $\erw{\theta_\lambda}$ of
the stochastic process above really solves the Black-Scholes equation one
again introduces the projection operator $F_\lambda$ and makes use of
the general theorem  (\ref{eq:real_expectation_formula}) to obtain
\begin{figure}
\begin{center}
\includegraphics[width=0.75\linewidth]{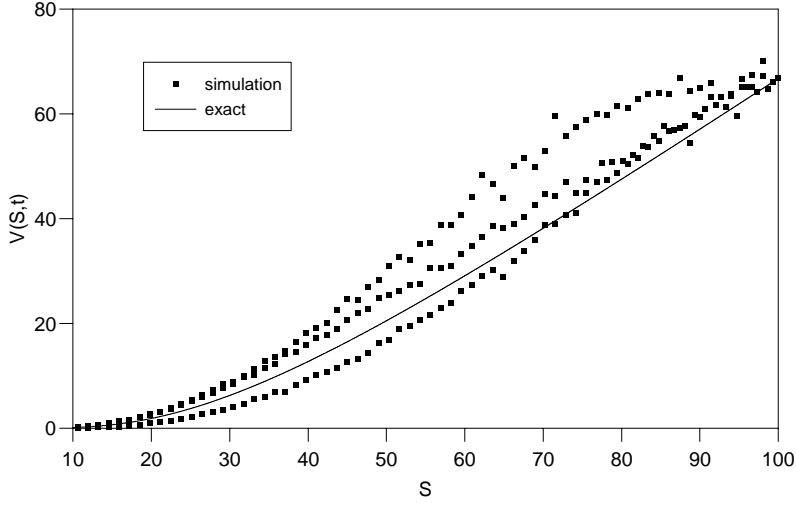}
\caption{Three exemplary realizations of the stochastic process from
  Eq.~(\ref{eq:fast_process}) (squares) and the corresponding exact
  solution (continuous line) of the Black--Scholes equation with
  Exercise price 60, $\sigma=0.2$, $r=0.06$ and  $T=10$. 
\label{fig:fig_6}}
\end{center}
\end{figure}

\begin{eqnarray}
\frac{\partial}{\partial t} \erw{\theta_\lambda} &=& \frac{
  \frac{2 \Delta_{\lambda-1}}{\Delta_{\lambda}+\Delta_{\lambda-1}}
 \erw{ \theta_{\lambda+1}} + 
  \frac{2 \Delta_\lambda}{\Delta_{\lambda}+\Delta_{\lambda-1}}
  \erw{\theta_{\lambda-1}}   - 2
  \erw{\theta_\lambda}}{\Delta_{\lambda-1} \Delta_\lambda}\nonumber\\
&+& (k-1) \frac{\erw{\theta_{\lambda+1}} -
  \erw{\theta_\lambda}}{\Delta_{\lambda+1}} - k \, \erw{\theta_\lambda}.
\end{eqnarray}
In the continuum limit the expectation value $\erw{\theta_\lambda}$ of
this stochastic process thus again solves the Black-Scholes equation, but
now on a non-uniform grid.
The transition probability $w(\theta, \tilde \theta)$ becomes
\begin{eqnarray}
w(\theta, \tilde \theta) &=& \frac{1}{\alpha_1} \sum_\mu 
\delta(\theta, r_\mu^+(\tilde \theta)) + \delta(\theta, r_\mu^-(\tilde \theta))
+ \frac{k-1}{\alpha_2} \sum_\mu \delta(\theta, s(\tilde
\theta))\nonumber \\ &&
+ \frac{k}{\alpha} \sum_\mu \delta(\theta, c(\tilde \theta)).
\end{eqnarray}
The parameters $\alpha_i$ are chosen such that the relative size of
the transitions is smaller than $1$, hence they fulfil $\alpha <
1$, $\alpha_2/\Delta_\mu < 1$ and $2 \alpha_1/\Delta^2_\mu < 1$ for
every possible transition $\mu$. The smallest $\alpha_i$, which  is
$\alpha_1$,  determines the scaling of the total transition probability
\begin{equation}
w= \frac{2 n}{\alpha_1} + (k-1) \frac{n}{\alpha_2} + \frac{k}{\alpha}.
\end{equation}
Since $\alpha_1 \sim O(\Delta^2_\mu)$ and the total transition
probability scales as $O(n/\alpha_1)$ the numerical effort to simulate
the stochastic process scales according to  $O(n/\Delta_\mu^2)$.
This is the same dependence on the
discretization as in the previous section with a stochastic process on
a uniform grid. But since now the grid can be chosen appropriate one
needs much less grid points and the algorithm is much faster.

\subsection{Fast stochastic process with non-uniform discretization}

\begin{figure}
\begin{center}
\includegraphics[width=0.75\linewidth]{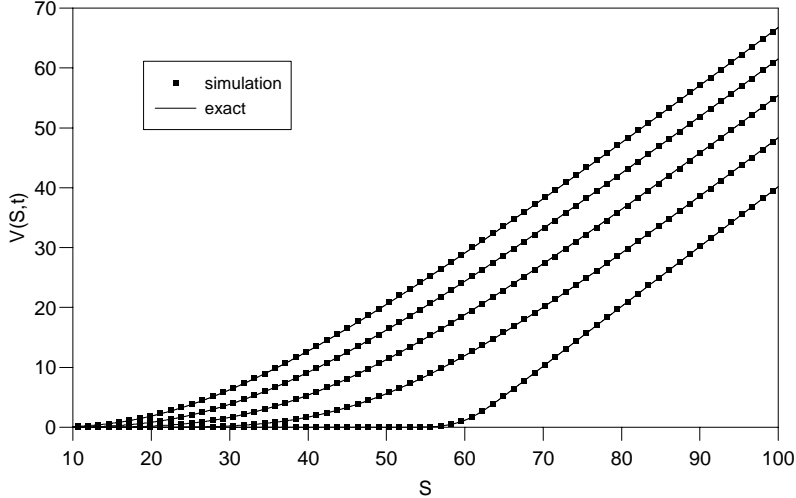}
\caption{Time evolution of the analytic solution (line) and results of a direct stochastic
  simulation (squares) according to Eq.~(\ref{eq:fast_process})
  averaged over 200 realizations of the
  Black-Scholes equation for an European call with
  Exercise price 60, $\sigma=0.2$, $r=0.06$ at the time points  $T=10, 7.5, 5,
  2.5, 0.05$.
\label{fig:fig_1}}
\end{center}
\end{figure}
The use of a non-uniform grid reduces the number of grid points to be
used in a numerical simulation. But there is still a
$O(n/\Delta_\mu^2) \approx O(1/\Delta_\mu^3)$ dependence of the total
transition rate which makes the algorithm slow. To get rid of this
dependence the stochastic process corresponding to the diffusive part
is again improved. In analogy to equilibrium Monte
Carlo simulation of lattice systems  \cite{binder.1995,landau.2000},
where one generates a new configuration of the whole lattice in a
single sweep, one defines
\begin{equation}
\ca{A}^{\text{sweep}}_{\text{diff}} =  \frac{1}{\alpha_1 } \left[
\frac{1}{\text{Det}(t^+)}  \ca{T}^+ +
\frac{1}{\text{Det}(t^-)}\ca{T}^-  -2 \, {\mathbb I} \right],
\end{equation}
with the map $t^\pm$ given by
\begin{equation}
t^\pm = \sum_\mu r_\mu^\pm.
\end{equation}
This process thus updates every $\theta_\lambda$ at once. The total
time evolution operator is now given by
\begin{equation}\label{eq:fast_process}
\ca{A}^{\text{sweep}}_{\text{BS}} = \ca{A}^{\text{sweep}}_{\text{diff}} +
\ca{A}^{\text{n.u.}}_{\text{conv}} + \ca{A}_{\text{chem}}.
\end{equation}
For the
transition probability one obtains
\begin{eqnarray}
\lefteqn{
w(\theta, \tilde \theta) = \frac{1}{\alpha_1} \;
\delta(\theta, t^+(\tilde \theta)) + \frac{1}{\alpha_1} \; \delta(\theta, t^-(\tilde \theta))}\nonumber\\
&+& \frac{k-1}{\alpha_2} \sum_\mu \delta(\theta, s(\tilde \theta))
+ \frac{k}{\alpha} \sum_\mu \delta(\theta, c(\tilde \theta))\;.
\end{eqnarray}
Since $\alpha_1$ again scales as $O(1/\Delta_\mu^2)$ the total
transition probability now also scales according to
$O(1/\Delta_\mu^2)$. This of course makes the algorithm significantly
faster.

\subsection{Analysis of the algorithm}

\begin{figure}
\begin{center}
\includegraphics[width=0.75\linewidth]{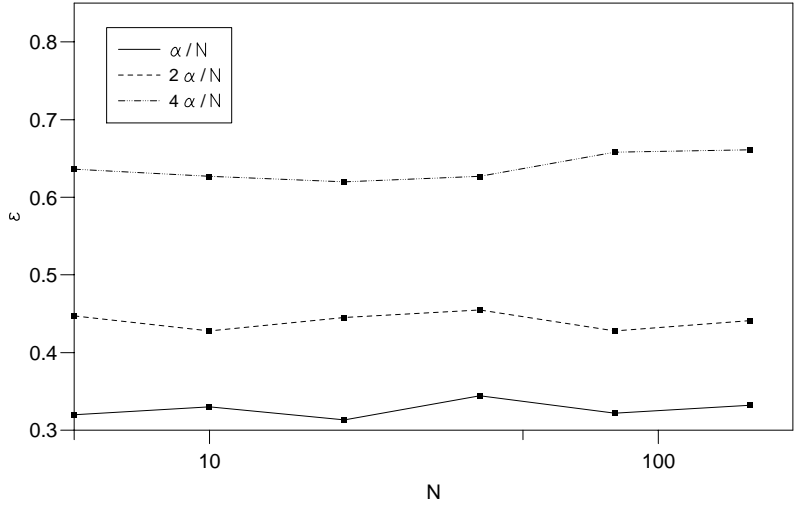}
\caption{Numerical root mean square error $\epsilon$ of the solution of the Black--Scholes
  equation for an European call with Exercise price 60, $\sigma=0.2$,
  $r=0.06$ according to the master equation (\ref{eq:fast_process})
  for three different ratios $\alpha/N$. The parameter values for the
  first simulation ($\alpha/N$)  are
  $\alpha_1/N=10^{-7}$, $\alpha_2/N=10^{-5}$ and $\alpha_3/N=10^{-4}$.
\label{fig:fig_4}}
\end{center}
\end{figure}
Fig.~\ref{fig:fig_6} shows three exemplary solutions of the stochastic
process defined in (\ref{eq:fast_process}) which corresponds to a
non-uniform discretisation of the dimensionless underlying $\tilde S$.
In addition to the expected random fluctuations around the exact solution,
single realizations also show systematic deviations from the exact
solution for a wide range of values of the underlying asset. These
systematic and random fluctuations can be traced back to the two
different types of stochastic processes entering Eq.~
(\ref{eq:fast_process}). The generators
$\ca{A}^{\text{n.u.}}_{\text{conv}}$ and  $\ca{A}_{\text{chem}}$
describe single jump processes, where the jump process changes only one option value in asset
time space, this of course causes the random
fluctuations. In contrast the systematic deviations stem from
$\ca{A}^{\text{sweep}}_{\text{diff}}$ which describes a jump process
changing all option values at once.\\
The numerical estimation of the option price from a sample of realizations
of the stochastic process is of course not affected by this behaviour
of single realizations. By averaging about several realizations the
estimated option price converges against the analytic solution, this
is shown in Fig.~\ref{fig:fig_1}.\\
Concerning the numerical root mean square error error
\begin{equation}
  \epsilon = \sqrt{\sum_k \left[ V(S_k,t) - \hat{V}(S_k,t) \right]^2},
\end{equation}
of the Monte Carlo
simulation of the Black--Scholes equations one has to distinguish two
types of parameters entering the numerical algorithm derived in the
previous section: the number of samples $N$ used to estimate the
solution $\hat{V}$ and the parameters $\alpha_i$ which control the size of the fluctuations. 
Figure \ref{fig:fig_4} shows the root mean squared error $\epsilon$ of a
Monte Carlo simulation for three different 
values of the ratio $\alpha_i/N$. Different simulations of the same
ratio were obtained  e.g. by decreasing $N$ and all three
parameters $\alpha_i$ by the same factor.
This figure clearly shows that the error only depends on this ratio
$\alpha_i/N$. In addition it can be seen that the error scales with
the square root of this ratio. This of course is a dependence which is
expected for a Monte Carlo simulation.\\
\begin{figure}
\begin{center}
\includegraphics[width=0.75\linewidth]{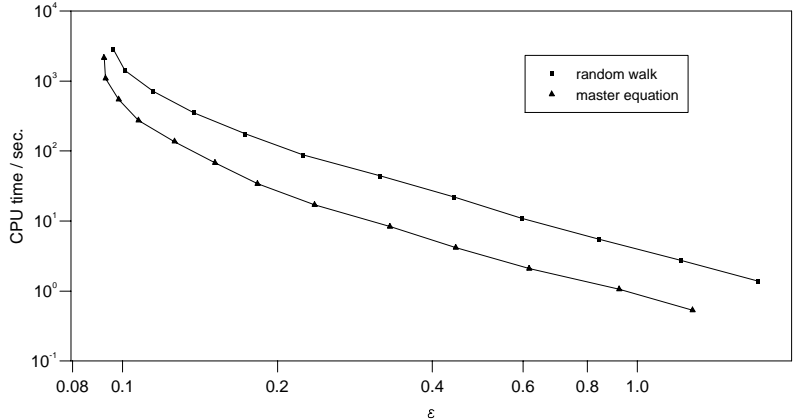}
\caption{ Comparison of the computation time in seconds needed for the pricing
  of an option by simulation of the stochastic dynamics of the
  underlying asset  and by a
  master equation formulation of the Black--Scholes equation. Shown is
  the computation time vs. the numerical root mean square error $\epsilon$.
\label{fig:fig_5}}
\end{center}
\end{figure}
In order to assess the numerical performance of the proposed Monte
Carlo method it is compared with the standard Monte Carlo approach
based on the simulation of the dynamics of the underlying asset. The numerical
error in a Monte Carlo solution has two sources. The systematic part
of the error stems from  the finite discretization of the stochastic
differential equation in the case of the simulation of the geometric
Brownian motion
and from the finite discretization of the state space when one solves
the master equation. In addition there is a random error from the
averaging over different solutions which decays in both cases as
$N^{-0.5}$ (if $\alpha$ is fixed in the master equation formulation).
Figure \ref{fig:fig_5} shows the time required to achieve a given precision
for the two methods in a parameter range where the systematic part of
the numerical errors are comparable. As can be easily seen the
solution of the master equation is significantly faster than the
standard approach.

\subsection{American options}\label{sec:american-options}

\begin{figure}
\begin{center}
\includegraphics[width=0.75\linewidth]{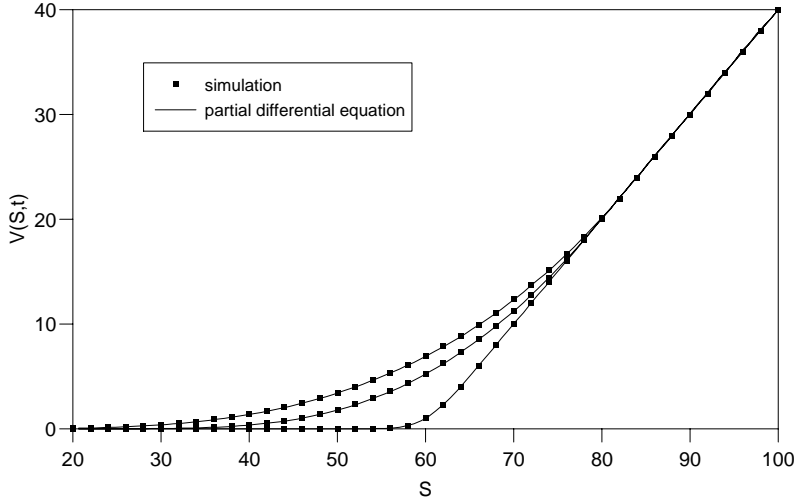}
\caption{Time evolution for an American call with
  Exercise price 60, interest rate 0.07 and dividend yield 0.10
  according to the stochastic simulation (squares) and the solution of the
  corresponding partial differential equation (continuous line) for
  the time points $T=10, 7.5, 5$.
\label{fig:fig_3}}
\end{center}
\end{figure}

Up to now only European options have been considered. How can the
concept of mesoscopic Monte Carlo simulations be generalised for American
options? A straightforward generalisation of the algorithm simulating
the time evolution of the underlying asset is quite involved, since the
early exercise condition has to be fulfilled. Whenever the option
price falls below the payoff function the option is exercised and the
option value is thus given by the payoff. Hence in order to assure
that the early exercise condition is not violated sometime in the future
a straightforward generalisation of this Monte Carlo simulation 
would have to keep track of all these points in asset time
space which would make the algorithm very ineffective. Of course, there
are more advanced techniques to generalise the Monte Carlo approach to
American options, see \cite{boyle1997} for an overview.\\
The proposed mesoscopic approach based on a master equation whose
expectation value solves the original Black--Scholes equation can
be generalised to the valuation of American options. Since the
time direction in the dimensionless Eq. (\ref{eq:bs_dl}) has been
reversed and the payoff function is used as initial condition the
generalisation to American options is straightforward. 
One just has to simulate one time step according to the master equation
(\ref{eq:fast_process}) and average about the number of samples used 
to estimate the option price. Wherever this option price is below the
payoff function it is replaced with the payoff and the next time step
is simulated. Hence one assures that the early exercise condition is
fulfilled.\\
For this approach to work it is critical that first
the samples are averaged and that the early exercise condition is
applied thereafter. 
Applying the early exercise condition before
averaging would introduce a bias towards higher option
prices. Every time the stochastic realization is by chance below the
payoff function the option value is increased. But since the option
value  is never decreased this way it is intuitively clear that one 
would obtain higher option prices.
Fig. \ref{fig:fig_3} shows the result of such a
simulation for an American call with exercise price 60, interest rate
0.07, volatility 0.2 and a continuous dividend yield of 0.10.

\section{Conclusions}

In real markets the elegance of the perfect hedge of the
Black--Scholes approach to option pricing is generally lost. Already a
very simple model of the dynamics of the underlying stock in terms of
piecewise deterministic processes shows that the option price itself
may be regarded as a stochastic variable. However the dynamics of the
expectation value of this stochastic option price is governed by the
Black--Scholes equation. This consideration is the motivation for the
master equation approach to option pricing suggested in this paper. \\
The essential point of this approach is that the 
Black--Scholes equation is interpreted as the macroscopic equation
of an underlying mesoscopic stochastic process for the stochastic
option price variable. By using PDPs one can then
easily proceed to a master equation formulation of the option pricing problem.
In contrast to this the usually used microscopic approach   describes the
dynamics of the underlying asset by a stochastic differential
equation.\\
The master equation formulation of the option pricing theory offers
several advantages over the standard approach.
This formulation provides a general setting in which also other kinds
of jump processes may be easily embedded, without altering the
character of the equations. One may also include additional stochastic
processes for the  volatility and the interest rate and then arrive
at stochastic volatility and interest rate models
\cite{hull1987,scott1987}. This of course would lead to different
hedging strategies. One possibility to extend the proposed approach
beyond the Black--Scholes equation, which is currently under
investigation, is to use a hedging strategy proposed in \cite{bouchaud1,bouchaud2}\\
In addition the master equation formulation also
allows for the use of advanced simulation algorithms. It was shown that
it is possible to construct a master equation whose transition
probability is constant in time and does not depend on the current
state of the system. As shown in section \ref{sec:general-theory} this
allows for fast numerical algorithms for the computation of the option price.\\
Using the standard Black--Scholes equation as an example we have 
demonstrated how to construct  such numerically  efficient
stochastic processes underlying the partial differential equation. As
can be seen from the derivation of the general theory in section
\ref{sec:general-theory}, it is clear that this approach is not
restricted to the standard Black--Scholes equation but
can be applied to  generalisations of the latter.\\
The master equation formulation of the Black--Scholes equation is
numerically about a factor of two faster than the standard Monte Carlo
approach. This is mainly due to the fact that the proposed mesoscopic
approach does not simulate sample trajectories of the underlying asset
which, after averaging, results in the option price for the initial asset
value at a given time. In contrast to this the mesoscopic approach works in the whole
asset time space and generates sample option prices
in the whole state space during one realization and is hence much
faster.\\ 
Another advantage of the master equation formulation of the
Black--Scholes equation is that it allows
for a straightforward generalisation to price American options in the
same framework by just comparing the average option price with the
payoff function. \\
Summarising, the proposed master equation approach is located in
between the standard Monte Carlo approach for the  simulation of  the underlying
asset and the finite difference solution of the partial differential
equation. This approach tries to combine the numerical efficiency of
the solution of the  partial differential equation with the
advantages of the Monte Carlo approach. The main advantage of the
Monte Carlo approach is that it can easily be applied to
price options using also advanced stochastic models for the underlying.\\
Concluding, the mesoscopic approach we have formulated has been exploited to
implement a master equation approach to option pricing. Of course,
it will be of great interest to investigate in further work if
features of real markets correspond to the additional freedom
to choose the size of the fluctuations contained in the proposed
master equation formulation. This will eventually lead to a better
understanding of option pricing in real markets.


\end{document}